\documentstyle[epsfig]{qcdparis}
\begin{document}
\pagestyle{plain}
\title{Direct and Resolved Pomeron}

\author{H.-G. Kohrs}

\affil{II. Institut f\"ur Theoretische Physik, \\
Universit\"at Hamburg, Germany}

\abstract{
  We investigate the effect of a direct pomeron coupling to quarks
  in DIS and photoproduction of jets. The direct pomeron coupling
  generates a point-like contribution to the diffractive part of
  the structure function $F_2$, which is analysed on the basis of
  the latest H1 data. Our model assumptions for the pomeron
  structure are consistent with the measured data.}

\resume{
  Nous \'etudions l'~effet du couplage direct du pomeron aux quarks
  pour des jets dans des \'ev\'enements de DIS et de photoproduction. Le
  couplage direct du pomeron engendre une contribution dite "point-like"
  a la contribution diffractive de la fonction de structure $F_2$, qui a
  \'et\'e analys\'ee avec les derniers r\'esultats de H1. Les
  hypoth\`eses de notre mod\`ele pour la structure du pomeron
  est en accord avec les donn\'ees mesur\'ees.}
\vspace{-10mm}
\twocolumn[\maketitle]

\fnm{7}{Talk given in the session of working group I$+$II
at the Workshop on Deep Inelastic Scattering and QCD,
Paris, April 1995}
\fnm{6}{Work done in Collaboration with B.A.~Kniehl and G.~Kramer}
\section{Introduction}
In diffractive production of hadronic final states in $ep$
scattering, the proton stays intact or becomes a low mass state.
Between the direction of the proton remnant, that goes down the
beam pipe, and the produced hadronic system there is no colour
flow, which allows of the possibility to observe large gaps in
rapidity between these directions. The experiments H1~\cite{H1}
and ZEUS~\cite{ZEUS} at HERA have measured the portion of diffractive
events to be $\approx 10\%$ of all events--not only in
photoproduction ($Q^2<0.01\,\mbox{GeV}^2$)
but also in DIS ($Q^2>10\,\mbox{GeV}^2$).

This paper is organized as follows. First, we describe our ansatz to
analyse the diffractive $ep$ scattering. In section 3, we consider the
pomeron structure function and the diffractive part of $F_2$. We find
consistency with the latest H1 data.

\section{Model for diffractive jet production}

There exist various phenomenological models \cite{IS,La,BH,In}
to describe the above mentioned diffractive
non--perturbative QCD phenomena quantitatively. We follow the model of
Ingelman and Schlein~\cite{IS}, where the proton splits off a
colourless object called pomeron (${I\hspace{-1mm}P}$),
which has quantum numbers of the vacuum. The ${I\hspace{-1mm}P}$
proton vertex is parametrized by a ${I\hspace{-1mm}P}$--flux
factor, that has been determined earlier in $p\overline{p}$ scattering
and can be used as input to analyse the HERA data.
It takes the form
\begin{eqnarray}
\label{flux}
      G_{I\hspace{-1mm}P/p}(x_{I\hspace{-1mm}P}) &=&
      \int_{-\infty}^{t_2} dt\,
      f_{I\hspace{-1mm}P/p}(t,x_{I\hspace{-1mm}P})
      \quad ,\\
\label{fluxt}
      f_{I\hspace{-1mm}P/p}(t,x) &=&
      \frac{1}{\kappa x} (A\exp^{\alpha t} + B\exp^{\beta t})
\end{eqnarray}
with $t_2=-m_p^2\frac{x_{I\hspace{-1mm}P}^2}{1-x_{I\hspace{-1mm}P}}$
and certain values for $\kappa$, $A$, $\alpha$, $B$, $\beta$.
The momentum transfer $t=(p-p')^2$ from the initial proton to the final
proton remnant has been integrated out, since the proton remnant is not
tagged. We emphasize that recently published H1 data~\cite{H1}
confirms this factorization scheme. However, since the pomeron
might not be a real particle, there could be problems with the
interpretation of (\ref{flux}) and (\ref{fluxt}) \cite{La}.

Surely, the pomeron is in some sense only a generic object, that serves
to parametrize a non perturbative QCD effect. Although it is not
considered as a physical particle that could be produced in the
s--channel, we employ the concept of structure functions for it.

For the hadron--like part of the unknown pomeron structure functions,
$G_{b/I\hspace{-1mm}P}$, we propose the ansatz
\begin{eqnarray}
      x\,G_{u/I\hspace{-1mm}P}(x)
      \hspace{-2mm}&=&\hspace{-2mm}
      x\,G_{\overline{u}/I\hspace{-1mm}P}(x)
      =x\,G_{d/I\hspace{-1mm}P}(x)
      =x\,G_{\overline{d}/I\hspace{-1mm}P}(x) \nonumber\\
      \hspace{-2mm}&=&\hspace{-2mm}
      6x(1-x) \frac{1}{5} \frac{1}{4} \quad,\nonumber\\
      x\,G_{s/I\hspace{-1mm}P}(x)
      \hspace{-2mm}&=&\hspace{-2mm}
      x\,G_{\overline{s}/I\hspace{-1mm}P}(x)
      =\frac{1}{2}x\,G_{u/I\hspace{-1mm}P}(x) \quad,\\
      x\,G_{g/I\hspace{-1mm}P}(x)
      \hspace{-2mm}&=&\hspace{-2mm}
      6x(1-x) \frac{3}{4}
      \nonumber
\end{eqnarray}
and vanishing charm contributions. These functions obey the sum
rule $\sum_b\int_0^1 dx\,x\,G_{b/I\hspace{-1mm}P}=1$ and are
defined for an input scale of $Q_0^2=2.25\,\mbox{GeV}^2$. We
carry out the usual DGLAP evolution to get the right $Q^2$
dependence of these functions~\cite{Vo}.

Besides the resolved contributions, the pomeron could couple
to quarks directly~\cite{BS,Br,CHPWW}.
Then, similarly to $\gamma\gamma$ scattering,
the $\gamma I\hspace{-1mm}P \rightarrow q\overline{q}$ cross
section also contributes to the pomeron structure function. Here,
we assume a direct vector coupling of the pomeron to the quarks
with coupling strength $c=1$. This is not really justified with
respect to the $C$ parity, but the $Q^2$ dependence, that is $\sim
\log Q^2$ at low $x$, is only weakly dependent on the spin structure
as one can see, for example, if one replaces the vector coupling
with a scalar one. To remove the collinear singularity, we introduce
the regulator quark masses $m_q$ and obtain for the point--like (pl)
part
\begin{eqnarray}
    \hspace{-2mm}&&\hspace{-2mm}
    \hspace{-5mm} x\,G_{q/I\hspace{-1mm}P}^{pl}(x,Q^2) \nonumber\\
    \hspace{-2mm}&&\hspace{-8mm}
    =\,x\,\frac{N_c}{8\pi^2}\,c^2 \Bigg\{\beta
    \bigg[-1+8x\left(1-x\right)-\frac{4m_q^2}{Q^2}x\left(1-x\right)
    \bigg] \nonumber\\
&&
    \hspace{-8mm} +\bigg[x^2+\left(1-x\right)^2
    +\frac{4m_q^2}{Q^2}x\left(1-3x\right)
            -\frac{8m_q^4}{Q^4}x^2 \bigg]
    \ln\frac{1+\beta}{1-\beta}\Bigg\}
    \nonumber
\end{eqnarray}
\begin{eqnarray}
\mbox{with} \hspace{15mm}
    \beta \hspace{-2mm}&=&\hspace{-2mm}
    \sqrt{1-\frac{4m_q^2x}{Q^2(1-x)}} \quad.
\end{eqnarray}
\ffig{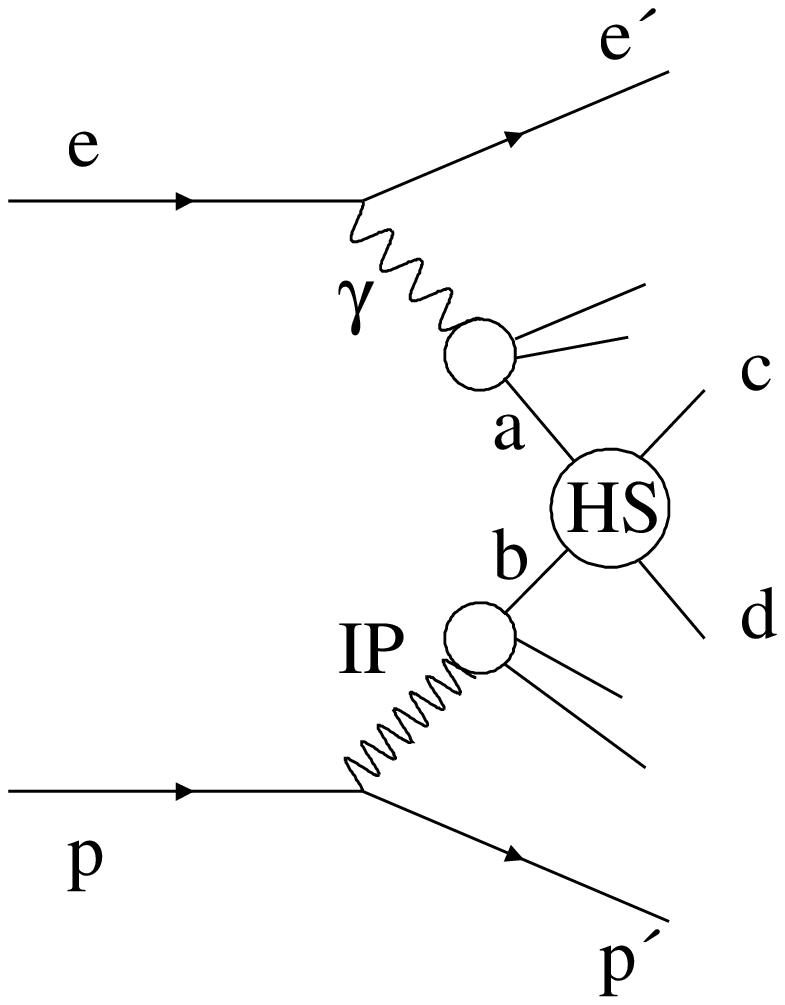}{30mm}{\em Generic diagram for the diffractive $ep$
scattering process with a resolved photon $\gamma$, a resolved
pomeron ${I\hspace{-1mm}P}$ and the hard subprocess $HS$.}{fig0}
\ffig{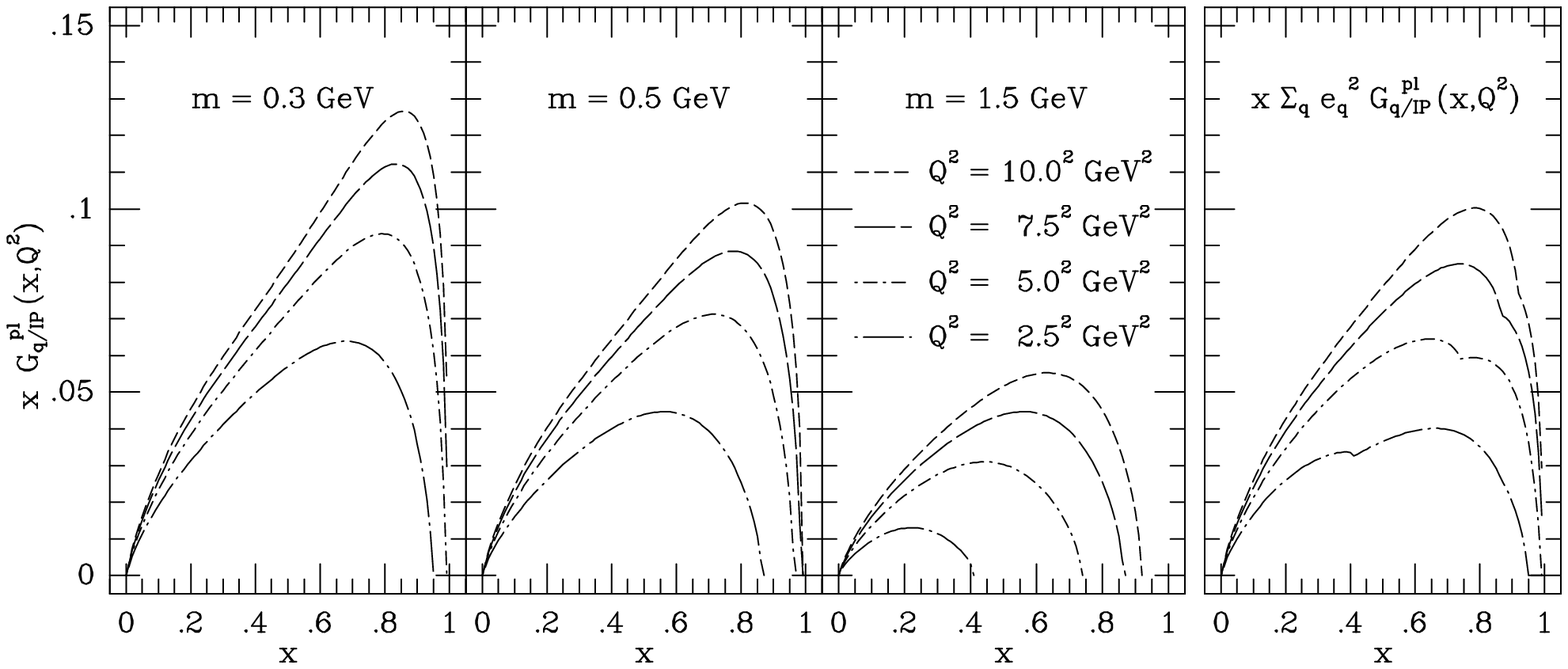}{32mm}{\em The pl distribution function for various
$Q^2$ values and our choice of regulator quark masses. The right
picture shows the $e_q^2$ weighted sum that enters into
$F_2^D$.}{fig1}
In \fref{fig1}, we have plotted the point--like
contribution for the three quark masses $m_u=m_d=0.3\,\mbox{GeV}$,
$m_s=0.5\,\mbox{GeV}$ and $m_c=1.5\,\mbox{GeV}$. The most right
picture shows the $e_q^2$ weighted sum of the pl contributions from
$u$, $d$, $s$ and $c$ quark to $F_2^D$. The dents in the curves
are caused by the charm thresholds. The point--like contribution
has its maximum in the upper $x$ region and increases
logarithmically with $Q^2$.

The calculation of the differential jet cross section
$\frac{d^2\sigma}{dy dp_T}$ of the process depicted in
\fref{fig0} is straight forward. In the case of
photoproduction, we have the usual factorization of the photon
flux factor at the electron vertex~\cite{WW}. As is well known
in photoproduction, the photon can be resolved or couples directly
to the final state quarks. For the resolved photon structure
functions, we take the parametrizations of GRV~\cite{GRV}.
Results and a detailed discussion of the one and two jet cross
sections can be found in \cite{KKK}. In DIS, the photon is always
direct. Jet cross sections in the $\gamma^*p$ system have been
carried out. However, at the moment, absolute experimental data for
rapidity distributions or $p_T$ spectra are not available for us.
A comparison with our predictions on diffractive jet production in
DIS has to be made in the future.

\section{The diffractive contribution to $F_2$}

To leading order in $\alpha_s$, only the quark distribution of the
pomeron enters into the deep--inelastic ${I\hspace{-1mm}P}$
structure function $F_2^{I\hspace{-1mm}P}(\beta,Q^2)$, which is
\begin{eqnarray}
\label{F2IP}
    \hspace{-2mm}&&\hspace{-2mm}
    \hspace{-7mm} F_2^{{I\hspace{-1mm}P}}(\beta,Q^2) \\
      \hspace{-2mm}&=&\hspace{-2mm}
    \sum_q
    e_q^2\,\beta\,\left[G_{q/{I\hspace{-1mm}P}}(\beta)
    + G_{\overline{q}/{I\hspace{-1mm}P}}(\beta)
    + 2\,G_{q/{I\hspace{-1mm}P}}^{pl}(\beta,Q^2)\right]
    \nonumber\quad.
\end{eqnarray}
The comparison with preliminary 1993 H1 data \cite{H1} is shown
in  \fref{fig2}. If factorization holds, which is favoured by
this experiment for a large range of $\beta$ and $Q^2$ values,
the data points are proportional to
$F_2^{{I\hspace{-1mm}P}}(\beta,Q^2)$, i.e.
$\tilde{F}_2(\beta,Q^2)=k\,F_2^{{I\hspace{-1mm}P}}(\beta,Q^2)$
with a constant $k$ that is in principle determined by the
${I\hspace{-1mm}P}$-flux factor.
To check the $Q^2$ evolution and shapes, we do not need the exact
value of $k$, so, we choose an appropriate value for $k$
and find that our model fits the shape of the data curves well.
Especially, the $Q^2$ evolution is fitted better for a combined
ansatz, i.e., quarks in the pomeron {\it and} pl part (solid
curves in \fref{fig2}), than for the DGLAP evolved quark
distribution (short dashed line in \fref{fig2}) or pl part alone.
An alternative possibility to fit the data has been represented in
\cite{ZEUS}.
\ffig{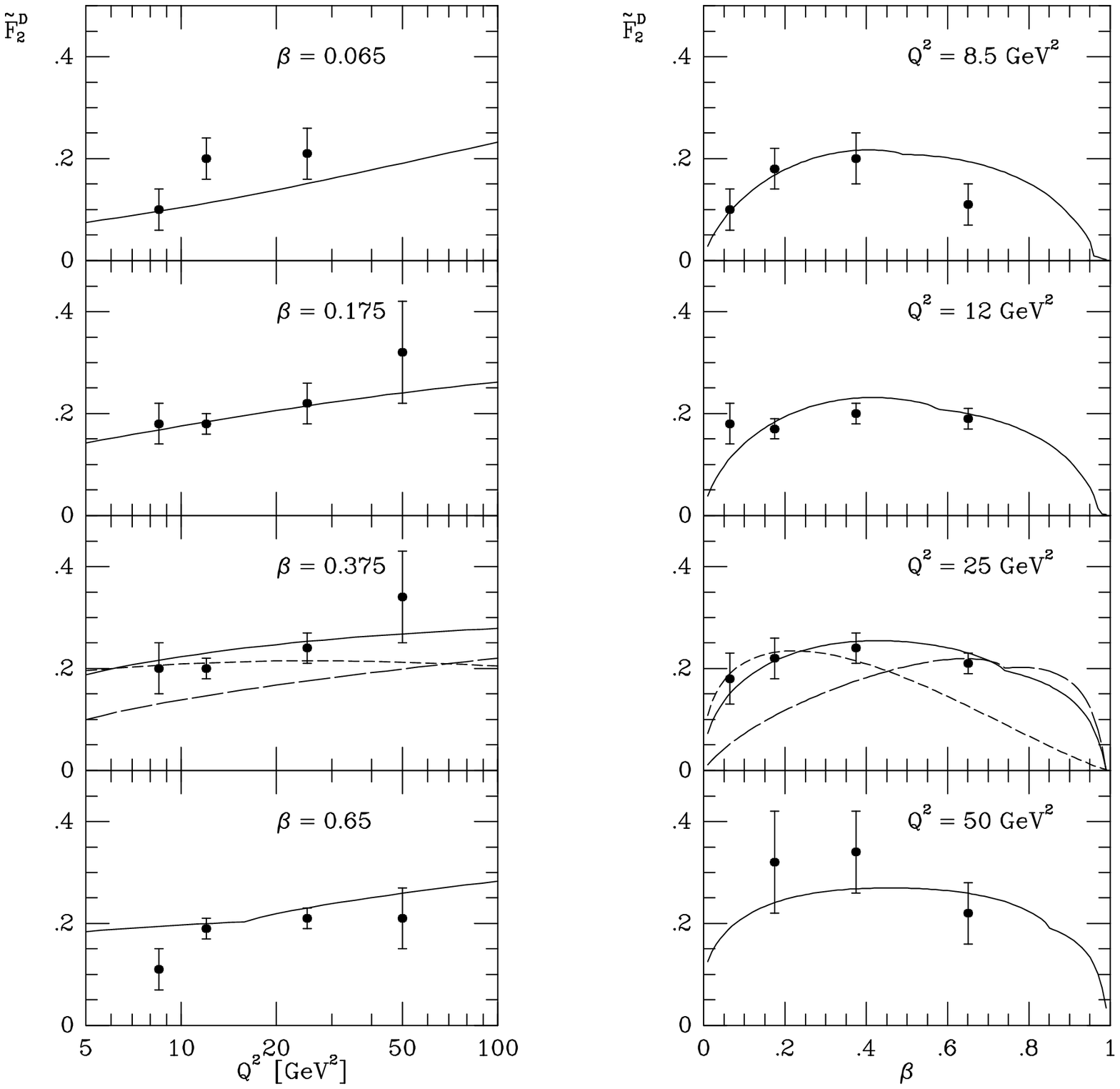}{83mm}{\em Comparison of our model predictions with
preliminary H1 data. An important feature of the point--like
contribution (long--dashed line, rescaled) is the filling up of
the quark distributions (short--dashed line) at higher $\beta$
where they become less dominant. This results in flatter curves
for the combined distribution (full line) and a better fit to
the shape of the data.}{fig2}

Finally, we fold the pomeron structure function (\ref{F2IP}) with the
pomeron flux factor in eq.~(\ref{flux}) to get the diffractive part
of the proton deep--inelastic structure function $F_2^D(x,Q^2)$.
The relation is
\begin{eqnarray}
\label{F2D}
    \hspace{-2mm}&&\hspace{-2mm}
    \hspace{-16mm} F_2^{D}(x,Q^2) \\
    \hspace{-2mm}&=&\hspace{-2mm}
    \int_x^{x_0}
    dx_{I\hspace{-1mm}P}\,\int_{t_1}^{t_2} dt\,
    f_{{I\hspace{-1mm}P}/p}(x_{I\hspace{-1mm}P},t)
    F_2^{I\hspace{-1mm}P}(\beta,Q^2) \nonumber
\end{eqnarray}
with $\beta=x/x_{I\hspace{-1mm}P}$. In \fref{fig3}, we compare
with 1993 H1 data~\cite{H1}. Again, we emphasize the consistency of the
choice $c=1$ for the direct pomeron coupling in our model with the data.
We use this value in the calculation of the diffractive jet cross
sections.
\ffig{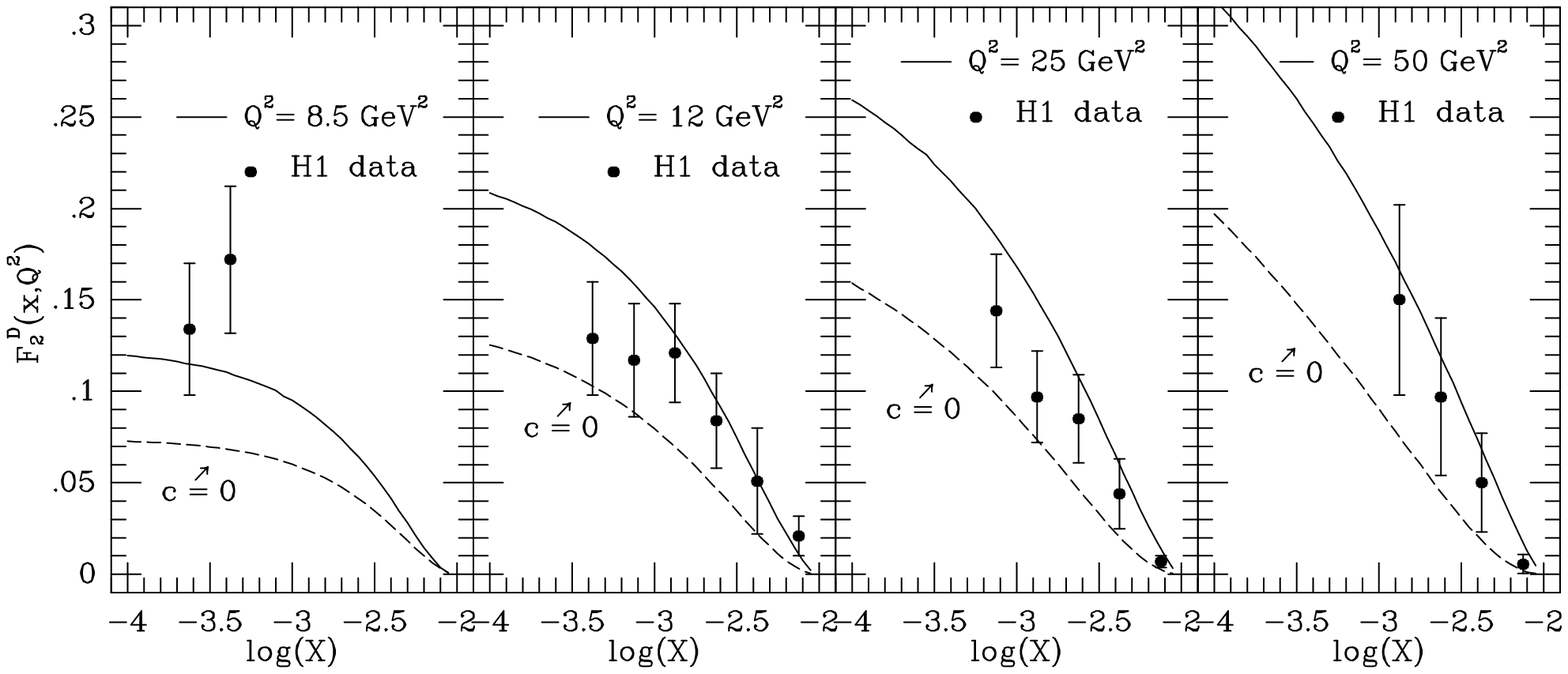}{32mm}{\em $F_2^D(x,Q^2)$ compared to
preliminary 1993 H1 data. The dashed lines represent
only the contributions from quarks in the resolved pomeron,
while for the solid curves the pl contribution with $c=1$
is included.}{fig3}
\section{Conclusion}
In summary, we have studied the effect of a direct pomeron coupling
on diffractive jet production at HERA. The concept of pomeron
structure functions with DGLAP $Q^2$ evolution has to be enlarged
if the pomeron has a direct coupling to quarks. We have included
the additional point--like part in the analysis of diffractive
$F_2$ data and find consistency. However, our analysis
is model dependent and second, we have compared only the shape
and not the normalization.
\begin{center}
{\large\bf Acknowledgements}
\end{center}
It is a pleasure to thank the conveners of the working group I$+$II
(Structures, Diffractive Interactions and Hadronic Final States)
and the organizing committee.
%


\Bibliography{100}
%
\bibitem{H1}
T. Ahmed et al., H1 Collaboration, Preprint~DESY~95--006;
Preprint~DESY~95--036 and Phys.~Lett.~B348(1995)681.

\bibitem{ZEUS}
M. Derrick et al., ZEUS Collaboration, Preprint~DESY~95--093.

\bibitem{IS}
G. Ingelman and P.E. Schlein, Phys.~Lett.~B152(1985)256.

\bibitem{La}
P.V. Landshoff, talks at Photon '95 (Sheffield) and DIS'95 (Paris).

\bibitem{BH}
W. Buchm\"uller, talk at DIS'95 (Paris); \\
W. Buchm\"uller and A. Hebecker, Preprint DESY~95--077.

\bibitem{In}
G. Ingelman, talk at DIS'95 (Paris).

\bibitem{Vo}
A. Vogt, private communication.

\bibitem{BS}
A. Berera and D.E. Soper, Phys.~Rev.~D50(1994)4328.

\bibitem{Br}
A. Brandt et al., UA8 Collaboration, Phys. Lett. B297(1992)417.

\bibitem{CHPWW}
J. C. Collins, J. Huston, J. Pumplin, H. Weerts and J.J. Whitmore,
Phys.~Rev.~D51(1995)3182.

\bibitem{WW}
C.F.v. Weizs\"acker, Z.~Phys.~88(1934)612;
E.J.~Williams, Phys.~Rev~45(1934)729.

\bibitem{GRV}
M. Gl\"uck, E. Reya and A. Vogt, Phys.~Rev.~D45(1992)3986;
Phys.~Rev.~D46(1992)1973.

\bibitem{KKK}
B.A. Kniehl, H.-G. Kohrs and G. Kramer, Preprint DESY~94--140 and
Z.~Phys.~C65(1995)657.

\end{thebibliography}
\end{document}